\documentclass{article}

\usepackage{arxiv}

\usepackage[utf8]{inputenc} 
\usepackage[T1]{fontenc}    
\usepackage{hyperref}       
\usepackage{url}            
\usepackage{booktabs}       
\usepackage{amsfonts}       
\usepackage{nicefrac}       
\usepackage{microtype}      

\usepackage{graphicx} 

\usepackage{amsbsy}
\usepackage{amsfonts}
\usepackage{amsmath}

\providecommand{\bo}{\mathbf}

\providecommand{\bs}{\boldsymbol}
\providecommand{\cov}{\mathrm{\bo COV}}
\providecommand{\E}{\mathrm{\bo E}}

\providecommand{\FOBI}{\mathrm{FOBI}}
\providecommand{\JADE}{\mathrm{JADE}}
\providecommand{\kJADE}{\mathrm{kJADE}}
\providecommand{\DF}{\mathrm{DF}}
\providecommand{\ADF}{\mathrm{ADF}}
\providecommand{\SF}{\mathrm{SF}}
\providecommand{\SSF}{\mathrm{S2F}}
\providecommand{\diag}{\mathrm{diag}}
\providecommand{\tr}{\mathrm{tr}}

\providecommand{\TANH}{\mathrm{tanh}}
\providecommand{\GAUS}{\mathrm{gauss}}
\providecommand{\POW}{\mathrm{pow3}}
\providecommand{\clr}{\mathrm{clr}}
\providecommand{\ilr}{\mathrm{ilr}}

\newtheorem{definition}{Definition}

\newtheorem{result2}{Key result}

\title{Independent component analysis for compositional data}

\date{} 					

\author{Christoph~Muehlmann \\
	Institute of Statistics \& Mathematical Methods in Economics \\
	Vienna University of Technology \\
	\texttt{christoph.muehlmann@tuwien.ac.at} \\
	\And
	Kamila~Fa\v cevicov\'a \\
	Department of Mathematical Analysis and Applications of Mathematics \\ 
	Palack\'y University Olomouc \\
	\texttt{kamila.facevicova@gmail.com} \\
	\And
	Al\v zb\v eta Gardlo \\
	Department of Clinical Biochemistry \\ 
	University Hospital Olomouc and Palack\'y University Olomouc \\
	\texttt{alzbetagardlo@gmail.com} \\
	\And
	Hana~Jane\v ckov\'a \\
	Laboratory for Inherited Metabolic Disorders, Department of Clinical Biochemistry \\ 
	University Hospital Olomouc and Palack\'y University Olomouc \\
	\texttt{janeckovah@gmail.com} \\
	\And
	Klaus~Nordhausen \\
	Institute of Statistics \& Mathematical Methods in Economics \\
	Vienna University of Technology \\
	\texttt{klaus.nordhausen@tuwien.ac.at} \\
}

\renewcommand{\headeright}{}
\renewcommand{\undertitle}{}

\begin{document}

\maketitle

\begin{abstract}
Compositional data represent a specific family of multivariate data, where the information of interest is contained in the ratios between parts rather than in absolute values of single parts. The analysis of such specific data is challenging as the application of standard multivariate analysis tools on the raw observations can lead to spurious results. Hence, it is appropriate to apply certain transformations prior further analysis. One popular multivariate data analysis tool is independent component analysis. Independent component analysis aims to find statistically independent components in the data and as such might be seen as an extension to principal component analysis. In this paper we examine an approach of how to apply independent component analysis on compositional data by respecting the nature of the former and demonstrate the usefulness of this procedure on a metabolomic data set. 
\end{abstract}



\section{Introduction}
\label{sec:intro}

Independent component analysis (ICA) is a well-established data analysis method in signal processing with the goal of recovering hidden signals that are usually meant to have a physical meaning. In recent years ICA methods have attracted increasing interest in the statistics community as extension of normality based multivariate methods that only use second order moments. In principle, ICA can be seen as a refinement of principal component analysis where, after removing second order information, higher order moments are used to search for hidden structures which are not visible in the principal components. Classical ICA methods are mainly developed for independent and identical distributed observations in a Euclidean space. Nevertheless, these methods are also applied for example on time series, spatial data, etc., but to the best of our knowledge not on iid compositional data. 

Compositional data is special in the way that the entries (parts) of a $d$-variate vector are positive and carry relative rather than absolute information about the respective observation of interest. Moreover, the parts of the compositional vector are by nature not independent and in some specific situations, e.g. when all parts are bounded by a constant sum constraint, a spurious correlation between them is present. Therefore, compositional data lies on a simplex and does not follow the Euclidean geometry. Examples of compositional data are geochemical data where the chemical composition of soil samples is of interest, the composition of nutrients of food intake or the distribution of market shares. For further details and examples of compositional data see for example \cite{Aitchison1986,EgozcuePawlowskyGlahn2019,facevicova2016,FilzmoserHronTempl2018,MoraisThomasAgnanSimioni2018,pawlowsky2011,TrinhMoraisThomasAgnanSimioni2019}.

It is well established that standard multivariate methods should not be applied directly to compositional data. Either methods which take the geometry of compositional data into account or methods that transform compositional data in such a way that standard multivariate analysis tools can be applied are appropriate. In this paper we take the latter approach.

We review some basic ICA methods in Section~\ref{sec:ICA}. In Section~\ref{sec:compData} we describe compositional data and methods to transform such data into the real space. Based on the former two sections we present how ICA can be performed on compositional data in Section~\ref{sec:ICAcompData} and conclude the paper with the analysis of a metabolomic data set from healthy newborns in Section~\ref{sec:CaseStudy} and a discussion in Section~\ref{sec:Dis}.

\section{Independent component analysis}\label{sec:ICA}

From a statistical perspective, independent component analysis is usually formulated as a latent variable model as follows.

\begin{definition}
	An observable $p$-vector $\bo x$ follows the independent component (IC) model if
	\[
	\bo x = \bo A \bo z + \bo b,
	\] 
	where $\bo A$ is a $p \times p$ non-singular matrix, $\bo b$ a $p$-vector and the latent $p$-variate random vector $\bo z$ satisfies:
	\begin{description}
		\item[(A1)] $\E(\bo z)=\bo 0$ and $\cov(\bo z) = \bo I_p$,
		\item[(A2)] the components of $\bo z$ are independent, and
		\item[(A3)] at most one component of $\bo z$ is Gaussian.
	\end{description}
\end{definition}

Thus $\E(\bo x) = \bo b$ and $\cov(\bo x) = \bo A \bo A^\top$. The goal of ICA is to find a $p \times p$ matrix $\bo W$ such that $\bo W \bo x$ has independent components. Note however that in general it will not hold that $\bo W \bo x = \bo z$ as the IC model assumptions only fix the location and scale of $\bo z$ but not the signs or the order of the components. Therefore, for every solution $\bo W$ also $\bo P \bo J \bo W$ is a solution where $\bo P$ is a $p \times p$ permutation matrix (one $1$ per row and column, $0$ elsewhere) and $\bo J$ a $p \times p$ sign-change matrix (a diagonal matrix with $\pm 1$ on its diagonal). 

There are many suggestions in the literature how to estimate $\bo W$ based on a sample $\bo X = (\bo x_1,\ldots,\bo x_n)$ and for recent reviews see for example \cite{comon2010handbook,NordhausenOja2018}. Almost all ICA methods make, however, use of the following result:

\begin{result2}
	Let $\bo x$ follow the IC model and denote $\bo x^{st} = \cov(\bo x)^{-1/2}(\bo x-\E(\bo x))$, then there exists an orthogonal $p \times p$ matrix $\bo U$ such that
	\[
	\bo U^\top \bo x^{st} = \bo z.
	\]
\end{result2}

Thus, this result means that after estimating $\cov(\bo x)$ and $\E(\bo x)$ the problem is reduced from finding a general $p \times p$ matrix to a $p \times p$ orthogonal matrix. Also note that this means that the performance of ICA methods does not depend on the values of $\bo A$ and $\bo b$ as these are accounted for when standardizing the data. An unmixing matrix estimate is therefore obtained as $\bo W = \bo U^\top\cov(\bo x)^{-1/2}$ and different ICA approaches differ in the way they estimate $\bo U$. In the following we will show how some popular ICA methods estimate this rotation.  

\subsection{FOBI} \label{sec:FOBI}

Fourth order blind identification (FOBI) \cite{cardoso1989source} was one of the first ICA methods but is still popular as it has a closed form solution. For FOBI we need to define the scatter matrix of fourth order moments

{\small
\[ 
\cov_4(\bo x) = \frac{1}{p+2} \E\left((\bo x-\E(\bo x))^\top \cov(\bo x)^{-1} (\bo x-\E(\bo x)) (\bo x-\E(\bo x)) (\bo x-\E(\bo x))^\top \right).
\]}

Then we can define 
\begin{definition}
The FOBI unmixing matrix is $\bo W_\FOBI = \bo U_\FOBI^\top\cov(\bo x)^{-1/2}$ where the columns of $\bo U_\FOBI$ are given by the eigenvectors of $\cov_4(\bo x^{st})$.
\end{definition}

Denoting $\bo U_\FOBI  \bo D \bo U_\FOBI^\top$ the eigendecomposition of $\cov_4(\bo x^{st})$ needed to compute $\bo W_\FOBI$ it is obvious that FOBI is only unique when the eigenvalues contained in the diagonal matrix $\bo D$ are distinct. One can actually show that these eigenvalues are linked  to the kurtosis values of the independent components. For FOBI to be well-defined Assumption (A3) from the IC model needs to be replaced by the stronger assumption
\begin{description}
 \item[(A4)] The kurtosis values of the independent components 
 must be distinct.
 \end{description} 

FOBI is often the first ICA method applied as it is quick to compute, gives a fast first impression and its statistical properties are well known, see for example \cite{miettinen2014fourth,NordhausenVirta:2019} for more details. FOBI can also be of interest outside the IC model and can be seen as an invariant coordinate selection method \cite{TylerCritchleyDumbgenOja:2009}.

\subsection{JADE} \label{sec:JADE}

Assumption (A4) is considered highly restrictive. Joint approximate diagonalization of eigenmatrices (JADE) \cite{cardoso1993blind} can be seen as an extension of FOBI which relaxes this strict assumption.

For JADE we have to define the fourth order cumulant matrices
\[
\bo C_{ij}(\bo x) = \E\left(({\bo x^{st}}^\top \E_{ij} \bo x^{st}) \bo x^{st} {\bo x^{st}}^\top\right)- \bo E_{ij} - \bo E_{ij}^\top - \tr(\bo E_{ij})\bo I_p,
\] 
where $\bo E_{ij}= \bo e_i \bo e_j^\top$ with $\bo e_i$ being a vector of dimension $p$ with the $i$th element equals $1$ and 0 otherwise. As $i$ and $j$ range from 1 to $p$ there are in total $p^2$ such cumulant matrices. In the IC model $\bo C_{ij}(\bo z) = 0$ if $i\neq j$ and for the case where $i=j$ $\bo C_{ii}(\bo z)$ corresponds to the kurtosis of the $i$th component. 
The matrix of fourth moments can actually be expressed as
\[
\cov_4(\bo x) = \frac{1}{p+2} \sum_{i=1}^p \bo C_{ii} (\bo x) + (p+2) \bo I_p
,\]
meaning that it uses not all possible cumulant information. The idea of JADE is to exploit the information contained in all cumulant matrices.

\begin{definition}
The JADE unmixing matrix is $\bo W_\JADE = \bo U_\JADE^\top\cov(\bo x)^{-1/2}$ where $\bo U_\JADE$ is the maximizer of
\[
\sum_{i=1}^p \sum_{j=1}^p ||\diag(\bo U^\top \bo C_{ij}(\bo x^{st}) \bo U)||_F^2.
\]
\end{definition}
Thus, JADE tries to maximize the diagonal elements of $\bo U^\top \bo C_{ij}(\bo x^{st}) \bo U$ which is equivalent to minimize the off diagonal elements by the orthogonal invariance of the Frobenius norm $|| \cdot ||_F$. As in the IC model only $\bo C_{ii} (\bo z)$ is non-zero and corresponds to the kurtosis of $z_i$. This means that JADE relaxes the FOBI assumption (A4) to 
\begin{description}
 \item[(A5)] At most one independent component can have zero kurtosis.
 \end{description} 
 
For a finite sample joint diagonalization of more than two matrices needs to be carried out approximately, many algorithms that jointly diagonalize two or more matrices are available, see for example \cite{IllnerMiettinenFuchsTaskienNordhausenOjaTheis2015}. For the purpose of this paper we will use an algorithm based on Given's rotations \cite{clarkson1988remark}.

The statistical properties of JADE are for example given in \cite{miettinen2014fourth}, from an asymptotic point of view FOBI is never superior compared to JADE. JADE is however computationally more expensive, especially when the number of independent components grows as $p^2$ matrices need to be computed and jointly diagonalized.

Thus, as a compromise k-JADE was suggested in \cite{MiettinenNordhausenOjaTaskinen:2013}. The idea is to use not all matrices $\bo C_{ij}$ but only those whose indices are not too far apart, i.e. $|i-j|<k$. This requires however that the first step, the whitening step, is not done using just the covariance matrix but using $\bo W_\FOBI$.

\begin{definition}
Denote $\bo x^{st'} = \bo W_\FOBI(\bo x - \E(\bo x))$ and choose an integer $1\leq k\leq p$, then the k-JADE unmixing matrix is $\bo W_\kJADE = \bo U_\kJADE^\top\bo W_\FOBI$ where $\bo U_\kJADE$ is the maximizer of
\[
\sum_{|i-j|<k}^p  ||\diag(\bo U^\top \bo C_{ij}(\bo x^{st'}) \bo U)||_F^2.
\]
\end{definition}

The value $k$ is basically a tuning parameter, the intuition is that the multiplicities of the distinct non-zero kurtosis values of the independent components are at most $k$ and that there is at most one component having kurtosis zero. Usually $k$ is simply chosen by the user based on subjective knowledge. In \cite{VirtaLietzenIlmonenNordhausen2020} some guidelines for the selection are offered, which are however not very practical. The statistical properties of k-JADE are given in \cite{MiettinenNordhausenOjaTaskinen:2013, VirtaLietzenIlmonenNordhausen2020}. It can be shown that for a value of $k$ which fulfils the multiplicity condition, k-JADE is asymptotically as efficient as JADE but has, if $k$ is small, a much smaller computational complexity.

\subsection{FastICA} \label{sec:fastICA}

FOBI, JADE and k-JADE are often called algebraic ICA methods. Another large group of ICA methods is based on projection pursuit ideas, where the most prominent one is FastICA originally suggested in \cite{Hyvarinen99fastica}. Some of the many FastICA variants are discussed below.

The general idea of FastICA is to find the column vectors $\bo u_1,\ldots,\bo u_p$ of $\bo U$ which maximize the non-Gaussianity of the components of $\bo U^\top \bo x^{st}$. Non-Gaussianity of a univariate random variable $x$ is measured by $|E(G(x))|$ with some twice continuously differentiable and non-quadratic function $G$ that satisfies $E(G(y))=0$ for $y \sim N(0,1)$. The most popular choices for $G$ are
\begin{description}
  \item[$\POW$:] $G(x) = (x^4-3)/4$,
  \item[$\TANH$:] $G(x) = \log(\cosh(x))- c_t$ and
  \item[$\GAUS$:] $G(x) = -\exp(-x^2/2)- c_g$.
  \end{description}  

The constants $c_t = E(\log(\cosh(y))) \approx 0.375$ and $c_g = E(-\exp(-y^2/2)) \approx -0.707$ are normalizing constants. The derivatives of $G$, denoted $g$ are called non-linearities and are the name givers as 
$
\POW: \ g(x) = x^3, \ \TANH: \ g(x) = \tanh(x) \ \mbox{and} \ \GAUS: \ g(x) = x \exp(-x^2/2)
$.

\subsubsection{Deflation-based FastICA} \label{sec:DeffastICA}

FastICA was first suggested in \cite{HyvarinenOja:1997} using the non-linearity $\POW$ and finding the column vectors of $\bo U_\DF$ one after another which is now known as deflation-based FastICA.

\begin{definition}
The deflation-based FastICA unmixing matrix is defined as $\bo W_\DF = \bo U_\DF^\top \cov(\bo x)^{-1/2}$, where the $k$th column of $\bo U$, $\bo u_k$, maximizes
\[
|\E[G(\bo u_k^\top\bo x_{st})]|
\]
under the constraints $\bo u_k^T\bo u_k=1$ and $\bo u_j^T\bo u_k=0,\ j=1,\dots,k-1$.
\end{definition}

 To obtain estimates a modified Newton-Raphson algorithm is used  which iterates the following steps until convergence: 
\begin{align*}
&\bo u_k\leftarrow \E[g(\bo u_k^\top\bo x_{st})\bo x_{st}]-\E[g'(\bo u_k^\top\bo x_{st})]\bo u_k \\
&\bo u_k\leftarrow \left(\bo I_p-\sum_{l=1}^{k-1}\bo u_l\bo u_l^\top \right)\bo u_k \\
&\bo u_k\leftarrow ||\bo u_k||^{-1}\bo u_k.
\end{align*}
The last two steps perform the Gram-Schmidt orthonormalization.

The properties of deflation-based FastICA have been studied in detail in \cite{Ollila2010,nordhausen2011deflation}. One issue with deflation-based FastICA is that besides the global maximum it has many local maxima and the order in which the vectors $\bo u_k$ are found depends heavily on the initial value of the algorithm where in turn the estimation performance depends on the order in which the vectors $\bo u_k$ are found. Using asymptotic arguments, \cite{nordhausen2011deflation} suggested reloaded FastICA, which estimates first the independent components using FOBI or k-JADE and then derives an optimal order based on the estimated independent components.


The idea of reloaded FastICA to fix the extraction order based on asymptotic arguments was extended by \cite{MiettinenNordhausenOjaTaskinen:2014} to also select an optimal non-linearity for each component out of a candidate set of possible non-linearities. This is known as adaptive deflation-based FastICA. We will denote the adaptive deflation-based FastICA unmixing matrix as $\bo W_\ADF$. The candidate set of non-linearities suggested in \cite{MiettinenNordhausenOjaTaskinen:2014} contains for example the non-linearities presented in Table~\ref{NonLin_afica}.

\setlength{\tabcolsep}{0.5em}

\begin{table}[!ht]
\scriptsize
\centering
\caption{Table of default candidate set of nonlinearities of adaptive deflation-based FastICA, where $(x)_+=x$ if $x>0$ and 0 otherwise, and $(x)_-=x$ if $x<0$ and 0 otherwise. } 
\begin{tabular}{lll}

$g_1(x)=x^3$			   &  $g_6(x)=(x)_+^2+(x)_-^2$ &     $g_{11}(x)=(x-1.0)_+^2+(x+1.0)_-^2$ \\
$g_2(x)=\tanh(x)$        & $g_7(x)=(x-0.2)_+^2+(x+0.2)_-^2$ & $g_{12}(x)=(x-1.2)_+^2+(x+1.2)_-^2$ \\
$g_3(x)= x\exp(-x^2/2)$  & $g_8(x)=(x-0.4)_+^2+(x+0.4)_-^2$ & $g_{13}(x)=(x-1.4)_+^2+(x+1.4)_-^2$\\
$g_{4}(x)=(x+0.6)_-^2$     & $g_{9}(x)=(x-0.6)_+^2+(x+0.6)_-^2$ & $g_{14}(x)=(x-1.6)_+^2+(x+1.6)_-^2$\\
$g_{5}(x)=(x-0.6)_+^2$     & $g_{10}(x)=(x-0.8)_+^2+(x+0.8)_-^2$ & \\
\end{tabular}

\label{NonLin_afica}
\end{table}

\subsubsection{Symmetric FastICA} \label{sec:SymfastICA}

A FastICA variant estimating all directions in parallel was suggested in~\cite{Hyvarinen:1999}.

\begin{definition}
The  symmetric FastICA estimator $\bo W_\SF = \bo U_\SF^\top \cov(\bo x)^{-1/2}$ uses as criterion for $\bo U_\SF$
\[
\sum_{j=1}^p|\E[G(\bo u_j^\top\bo x_{st})]|
\]
which should be maximized under the orthogonality constraint $\bo U_\SF^\top \bo U_\SF= \bo I_p$.
\end{definition}

The steps of the iterative algorithm to compute $\bo U_\SF$ are 
\begin{align*}
&\bo u_k\leftarrow \E[g(\bo u_k^T\bo x_{st})\bo x_{st}]-\E[g'(\bo u_k^\top\bo x_{st})]\bo u_k,\ \ k=1,\dots,p \\
&\bo U_\SF^\top\leftarrow (\bo U_\SF^\top\bo U_\SF)^{-1/2}\bo U_\SF^\top.
\end{align*}

The first update step of the algorithm is similar to that of the deflation-based FastICA estimator. Whereas the orthogonalization step 
can be interpreted as taking an average over the vectors of the first step. This differs from the deflation-based approach where errors made in the $k$th direction carry on to the following directions and therefore the errors accumulate. This is often the reason why symmetric FastICA is usually considered superior to the deflation-based FastICA. However, there are also cases where the accumulation is preferable to the averaging. This 
occurs when some independent components are easier to find than the others. Statistical properties of symmetric FastICA are given in \cite{miettinen2014fourth,Wei2015,MiettinenNordhausenOjaTaskinenVirta:2017}.

\subsubsection{Squared symmetric FastICA} \label{sec:SqSymfastICA}

One of the most recent variants of FastICA is the squared symmetric FastICA estimator \cite{MiettinenNordhausenOjaTaskinenVirta:2017}. The idea of this estimator is to replace the absolute values in the objective function of the symmetric FastICA with squared values.

\begin{definition}
The squared symmetric FastICA estimator $\bo W_\SSF =$ $\bo U_\SSF^\top \cov(\bo x)^{-1/2}$ obtains $\bo U_\SSF$ as the maximizer of
\[
\sum_{j=1}^p(\E[G(\bo u_j^\top\bo x_{st})])^2
\]
under the orthogonality constraint $\bo U_\SSF^\top \bo U_\SSF=\bo I_p$.
\end{definition}

The steps of the resulting algorithm are
\begin{align*}
&\bo u_k\leftarrow \E[G(\bo u_k^\top\bo x_{st})](\E[g(\bo u_k^\top\bo x_{st})\bo x_{st}]-\E[g'(\bo u_k^\top\bo x_{st})]\bo u_k),\ \ k=1,\dots,p, \\
&\bo U_\SSF^\top\leftarrow (\bo U_\SSF^\top \bo U_\SSF)^{-1/2} \bo U_\SSF^\top.
\end{align*}

Thus, the first step of the algorithm equals the first step in the symmetric algorithm with an additional multiplication by $\E[G(\bo u_k^\top\bo x_{st})]$. Hence, the squared symmetric variant puts more weight on components that are ``more'' non-Gaussian, which most often, but not always, is advantageous. The properties of the squared symmetric FastICA estimator as well as comparisons to the deflation-based and symmetric FastICA method are given in \cite{MiettinenNordhausenOjaTaskinenVirta:2017}. \cite{MiettinenNordhausenOjaTaskinenVirta:2017} also show that if the non-linearity $\POW$ is used, symmetric squared FastICA is asymptotically equivalent to JADE.

Besides assumptions (A1)-(A3), deflation-based, symmetric and squared symmetric FastICA need further assumptions based on $G$ to ensure consistency. Assuming the order of the components is fixed as $|\E[G(z_1)]| \geq \dots \geq |\E[G(z_p)]|$, then it is required that for any $\bo z=(z_1,\dots,z_p)^\top$ with independent and standardized components and for any orthogonal matrix $\bo U=(\bo u_1,\dots,\bo u_p)$ the following holds:

For deflation-based FastICA
\begin{description}
	\item[(A6)] For all $k=1,\ldots,p$, $|\E[G(\bo u_k^\top\bo z)]|\leq |\E[G(z_k)]|$, when $\bo u_k^\top \bo e_j = 0$ for all  $j=1,\ldots,k-1$, where $\bo e_i$ is a $p$-vector with $i$th element one and others zero,
\end{description}

for symmetric FastICA
\begin{description}
	\item[(A7)] $|\E[G(\bo u_1^T\bo z)]|+\cdots+|\E[G(\bo u_p^T\bo z)]|  \leq \ |\E[G(z_1)]|+ \cdots+|\E[G(z_p)]| $
\end{description}

and for squared symmetric FastICA
\begin{description}
	\item[(A8)] $(\E[G(\bo u_1^T\bo z)])^2+\cdots+(\E[G(\bo u_p^T\bo z)])^2 
	 \leq \ (\E[G(z_1)])^2+ \cdots+(\E[G(z_p)])^2. $
\end{description}

It was proven for example in \cite{miettinen2014fourth} that $\POW$ fulfils these conditions but for example $\TANH$ and $\GAUS$ not for all possible source distributions.

From a computational point of view the advantage of both symmetric versions is, that the initial value of $\bo U$ is not important when the sample size is large, as the algorithms converge usually to the global maxima. 

To conclude this section we can point out, that FOBI, JADE, k-JADE, symmetric FastICA and squared symmetric FastICA are affine equivariant ICA methods which means that their performance does not depend on the mixing matrix. So from this point of view only deflation-based fastICA differs, which can be overcome when the reloaded version or adaptive version is used.
Affine equivariance will be of relevance later when applying the ICA methods to compositional data.

\section{Compositional data and its real space representation}\label{sec:compData}

A specific family of $d$-dimensional vectors is present when each entry (part) of a vector is positive and carries information about its contribution to the whole. In the following such multivariate observations are called (vector) compositional data, whose specifics were already described, utilized and analyzed in a wide range of applications \cite{pawlowsky2011}. The main property of compositional data is its relative nature, when the whole relevant information is contained in the ratios between parts rather than in the absolute values of the parts. Consider e.g. a vector describing a geochemical structure of soil, where each part represents the quantity of the given element in the sample. The quantity can be given either in absolute scale, like in mg of the component contained in the sample, or some of its relative alternatives, typically ppm. While the mg representation depends on the overall size of the sample, the ppm one does not, despite of that the ratios between parts remain unchanged and both representations are therefore from the compositional point of view equivalent.

Due to the relative nature of compositional data, the sample space of representations of a $d$-part compositional vector $\mathbf{x}$ forms a $d$-part simplex
$$
\mathcal{S}^d=\left\{\mathbf{x}=(x_1, \dots, x_d)^\top, \sum_{i=1}^d{x_i}=\kappa,  \kappa> 0  \right\},
$$
where the Aitchison geometry holds. The whole sample space is formed by equivalence classes of proportional vectors. Since most of the standard statistical methods are designed for real-valued data following the Euclidean geometrical structure, it is favourable to express a compositional vector in real coordinates prior to its analysis. One of the possible representations is the centered log-ratio (clr) transformation from $\mathcal{S}^d$ to $\mathbb{R}^d$ given by

$$
\mathrm{clr}(\mathbf{x})_i=\ln\frac{x_i}{g_m(\mathbf{x})}=\frac{1}{d}\sum_{j=1}^d{\ln\frac{x_i}{x_j}}, \quad \mathrm{for} \quad i=1, \dots, d,
$$
where $g_m(\mathbf{x})$ denotes the geometrical mean of all parts. The parts of the resulting clr vector can be interpreted in terms of dominance of the given compositional part within the whole composition or equivalently as its mean dominance over each part of the whole composition. The use of logarithm symmetrizes this relationship. Let us stress here, that the clr values depend on the set of compositional parts used for its computation and therefore the above interpretation holds true only when the whole composition is considered. Within the whole manuscript clr transformation based on all compositional parts will be of the interest. On the other hand, from its construction, the clr coefficients/variables are not linearly independent, sum up to zero, and therefore, the whole clr vector falls in a $(d-1)$-dimensional subspace of $\mathbb{R}^d$. This feature prevents from direct use of the clr representation within methods that require full rank data, like robust PCA \cite{filzmoser2009} or the above stated ICA methods. 

One possible workaround is the isometric log-ratio (ilr) transformation, which represents the compositional vector $\mathbf{x}$ in a system of $d-1$ orthonormal real coordinates. This system can be obtained directly from the clr vector as

$$
\mathrm{ilr}(\mathbf{x})=\mathbf{V}^\top\mathrm{clr}(\mathrm{x}),
$$
where the columns of the $d \times d-1$ log-contrast matrix $\mathbf{V}$ are given as $\mathbf{v}_i=\mathrm{clr}(\mathbf{\xi}_i)$ and the vectors $\mathbf{\xi}_i$, $i=1, \dots, d-1$ constitute an orthonormal basis in $\mathcal{S}^d$. See \cite{pawlowsky2011}, ch. 11 for details. 

The system of basis vectors $\{\mathbf{\xi}_1, \dots, \mathbf{\xi}_{d-1}\}$ is not uniquely given and can be chosen according to the purpose of further analysis. Since each system of ilr coordinates can be obtained as an orthogonal rotation of the others, its specific choice does not affect the results of their analysis, like e.g. predictions of the regression model with a compositional regressor or scores of the robust PCA model \cite{filzmoser2009,hron12}, but when it is needed, it can be selected according to an expert knowledge in order to obtain an optimal interpretation of the coordinates at hand \cite{egozcue05}. Since a specific interpretation of the ilr coordinates is not the main purpose here, the same system as in \cite{NordhausenOjaFilzmoserReimann2015} is used. The basis vectors $\mathbf{\xi}_i$ have the value $\exp\left(\sqrt{1/i(i+1)}\right)$ at the first $i$ positions, $\exp\left(-\sqrt{i/(i+1)}\right)$ at the position $i+1$ and $1$ at the remaining ones. Consequently, the columns of the log-contrast matrix are 
$$
\mathbf{v}_i=\sqrt{\frac{i}{i+1}}\left(\frac{1}{i}, \dots, \frac{1}{i}, -1, 0, \dots, 0 \right)^\top, \quad i=1, \dots, d-1 ~ .
$$
The ilr coordinates have the form of balances between the $i$-th part of the composition and all parts with lower indices
$$
\mathrm{ilr}(\mathbf{x})_i=\sqrt{\frac{i}{i+1}}\ln\left(\frac{(x_1\cdots x_i)^{1/i}}{x_{i+1}}\right), \quad \mathrm{for} \quad i=1, \dots, d-1. 
$$
Finally, the clr and ilr representations are mutually transferable through the contrast matrix $\mathbf{V}$
$$
\mathrm{clr}(\mathbf{x}) = \mathbf{V}\mathrm{ilr}(\mathbf{x})
$$
and also the back-transformation to the simplex is possible by using
$$
\mathbf{x}=\exp(\mathrm{clr}(\mathbf{x}))=\exp(\mathbf{V}\mathrm{ilr}(\mathbf{x})).
$$

\section{ICA for compositional data}
\label{sec:ICAcompData}
As described above, ICA is not reasonable for data following an Aitchison geometry, it is natural to transform the data first into the Euclidean space. However, as ICA methods start with whitening and therefore require full rank data, the ilr space is the most natural representation. Due to the affine equivariance property of the discussed ICA methods the particular used basis for the ilr transformation at most affects the order and signs of the estimated independent components. Hence, for compositional data ICA we have the following model assumption:
\[
\ilr(\bo x) = \bo A_{\ilr} \bo z + \bs b,
\]
where $\bo A_{\ilr}$ is a $(d-1) \times (d-1)$ full rank mixing matrix specific for a chosen ilr basis, $\bs b$ a $d-1$ dimensional location vector and $\bo z = (z_1,\ldots, z_{d-1})^\top$ a random vector with independent components, which are standardized so that $\E(\bo z) = \bo 0$ and $\cov(\bo z)= \bo I_{d-1}$.
When the unmixing matrix $\bo W_{\ilr}$ is estimated using one of the ICA methods described in Section \ref{sec:ICA}, the system of independent components is given by
\[
\bo z = \bo W_{\ilr} ( \ilr(\bo x) - \bs b ) = \bo W_{\ilr}  (\bo V^\top \clr(\bo x) - \bs b).
\]
As ilr coordinates are not directly related to the dominance of the original parts, the relationship between ilr and clr spaces can be exploited yielding a $(d-1) \times d$ ``clr'' loading matrix $\bo W_{\clr} = \bo W_{\ilr} \bo V^\top$, allowing interpretation of the independent components in the clr space. In the context of principal component analysis performed in the clr space, principal components can be understood as a new system of ilr coordinates \cite{pawlowsky-glahn11}. This is not the case for ICA, as the unmixing matrix $\bo W_{\ilr}$ (and consequently also  $\bo W_{\clr}$) is generally not restricted to be orthogonal.
Even if one does not believe in an independent component model, ICA transformations remain affine equivariant which means that $\bo z$ can be seen as intrinsic data representation with a coordinate system whose components are as independent as possible.

After performing ICA one is usually interested in either using $\bo z$ itself for further analysis such as classification, outlier identification, etc., with possible interpretation in ilr or clr space using the former defined loading matrices $\bo W_{\ilr}$ or $\bo W_{\clr}$, or, using ICA for noise or artifact removal. For that purpose the components of $\bo z$ are divided into a signal part $\bo z_s$ and a noise/artifact part $\bo z_n$. This defines also the partition of the unmixing matrix $\bo W_{\ilr}$ into $\bo W_{\ilr}^s$ and $\bo W_{\ilr}^n$ and the mixing matrix $\bo A_{\ilr} =\left(\bo W_{\ilr}\right)^{-1}$ into $\bo A_{\ilr}^s$ and $\bo A_{\ilr}^n$. $\bo A_{\ilr}^s$ is formed only by those columns of $\bo A_{\ilr}$, that correspond to the signal components $\bo z_s$. 
The pure signal can then be restored in the ilr, clr and original space by using 
\[
\ilr(\bo x)_s = \bo A_{\ilr}^s \bo z_s + \bo b , \quad 
\clr(\bo x)_s = \bo V\bo A_{\ilr}^s \bo z_s + \bo b, \quad \mathrm{and} \quad
\mathbf{x}_s=\exp(\bo V\bo A_{\ilr}^s \bo z_s + \bo b).
\]


\section{A case study in metabolomics}\label{sec:CaseStudy}
In order to demonstrate the above described methods, the data from a neonatal screening program in the Czech Republic was analyzed. Anonymous data were obtained from a retrospective study approved by the Ethics Committee of the University Hospital Olomouc which was part of a larger international study described in \cite{Fleischman2013}. Newborn screening is a preventive program that allows for an early detection of a selected spectrum of inborn metabolic diseases. At an age of 48-72 hours after birth several drops of blood from the heel of the child were sampled on a special paper and sent for analysis to the screening laboratory. The data at hand were constituted by the metabolite profile of over $10\thinspace 000$ healthy newborns, for each neonate the values of 48 metabolites were measured. Moreover, information about sex and birth weight was available. More specifically, the birth weight ranged between 300 and $5\thinspace 570$ grams and for newborns with very low birth weight (less than 1500 grams) a different metabolite structure can be expected, due to their prematurity and the artificial nutrition they receive. One of the main goals of metabolomics is to investigate interactions between metabolites, their dynamic changes and responses to stimuli. Biofluids, e.g. blood or urine, and also tissues are used for the analysis. On the one hand, the most frequently used approach for the data analysis is done through comparison of absolute values of biomarkers and reference ranges (data from the healthy population). On the other hand, the new trend of data evaluation is based on the
use of ratios of metabolite data. Relative changes are more relevant/informative than absolute values in diagnostics based on profiling. Therefore, metabolomic data can be considered as observations carrying relative information, i.e. as compositional data \cite{kalivodova2015}, and as such the above discussed methods can be applied. 

\begin{figure}[h]
    \begin{minipage}[t]{0.49\textwidth}
        \centering
        \includegraphics[width=0.98\textwidth]{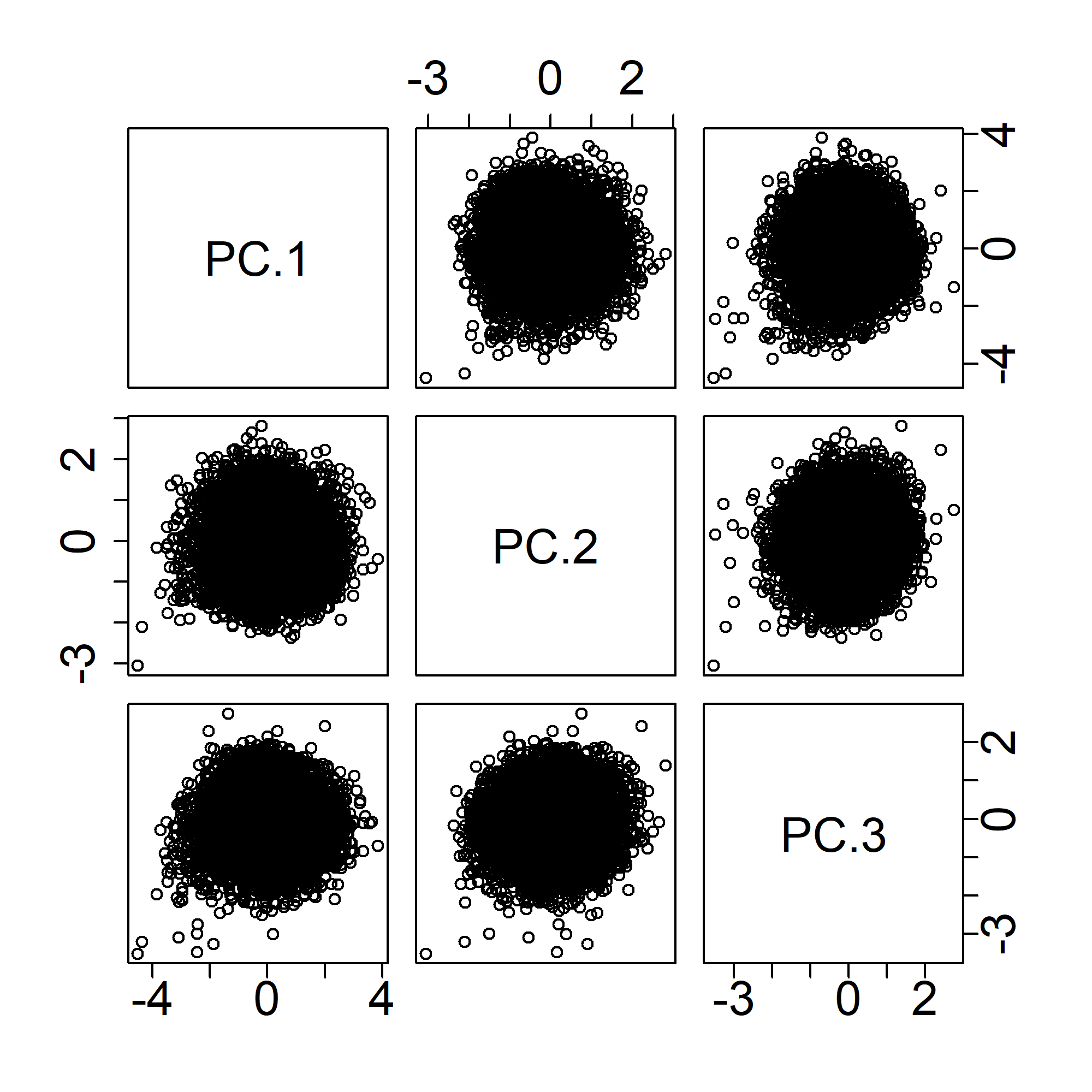}
    \end{minipage}
    \hfill
    \begin{minipage}[t]{0.49\textwidth}
        \centering 
        \includegraphics[width=0.9\textwidth]{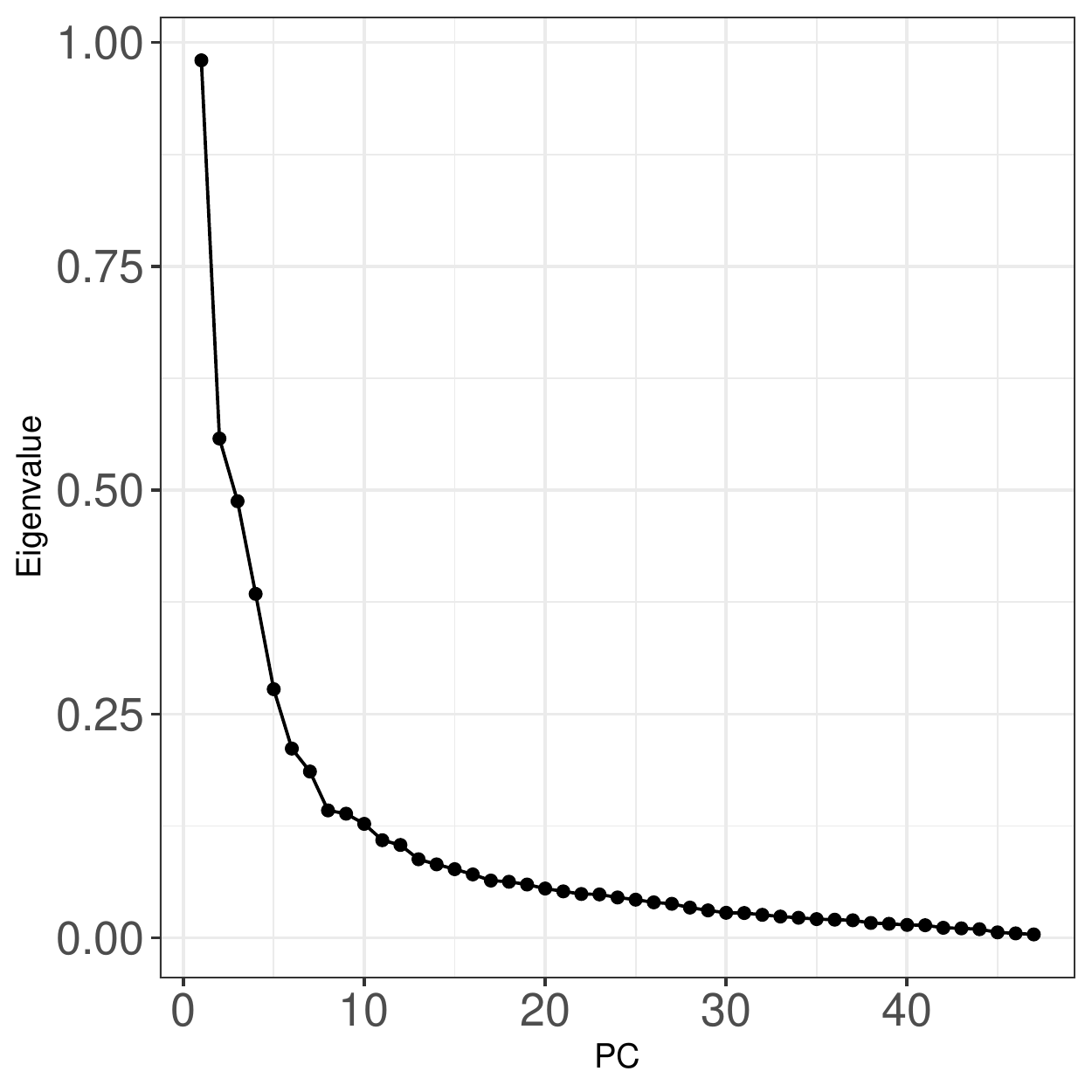}
    \end{minipage}
    \caption{Scatterplots of the first three principal components resulting from the compositional PCA (left) and scree plot of the respective explained variability (right).}
   \label{fig:pca}
\end{figure}

The following analysis was carried out in R 3.6.1 \cite{r_language} with the help of the packages JADE \cite{JADE_package}, fICA \cite{fICA_package}, compositions \cite{compositions_package} and robCompositions \cite{robCompositions_package}. As a first step, standard principal component analysis (PCA) was performed on clr transformed data. There were no significant patterns visible within the first three principal components, see Figure \ref{fig:pca}, left. The whole dataset form one quite compact cluster with no outliers. Moreover, the variance explained by the first components is low (around 20 \% for the first PC) (Figure \ref{fig:pca}, right) and therefore PCA does not seem to deal well with the issue of outlier detection, grouping as well as dimension reduction in that case.

As PCA seems not to reveal any clear structure we applied FOBI, k-JADE, with $k = 5$, and adaptive deflation-based FastICA to the ilr representation of the data (the dimension $p = 47$ was already too large for JADE). For easier comparison the components from all three ICA methods were ordered according to their kurtosis values. As all three ICA methods showed similar results, we focus in our presentation and discussion of the components on those from adaptive deflation-based FastICA. 



\begin{figure}
    \begin{minipage}[t]{0.49\textwidth}
        \centering
        \includegraphics[width=0.98\textwidth]{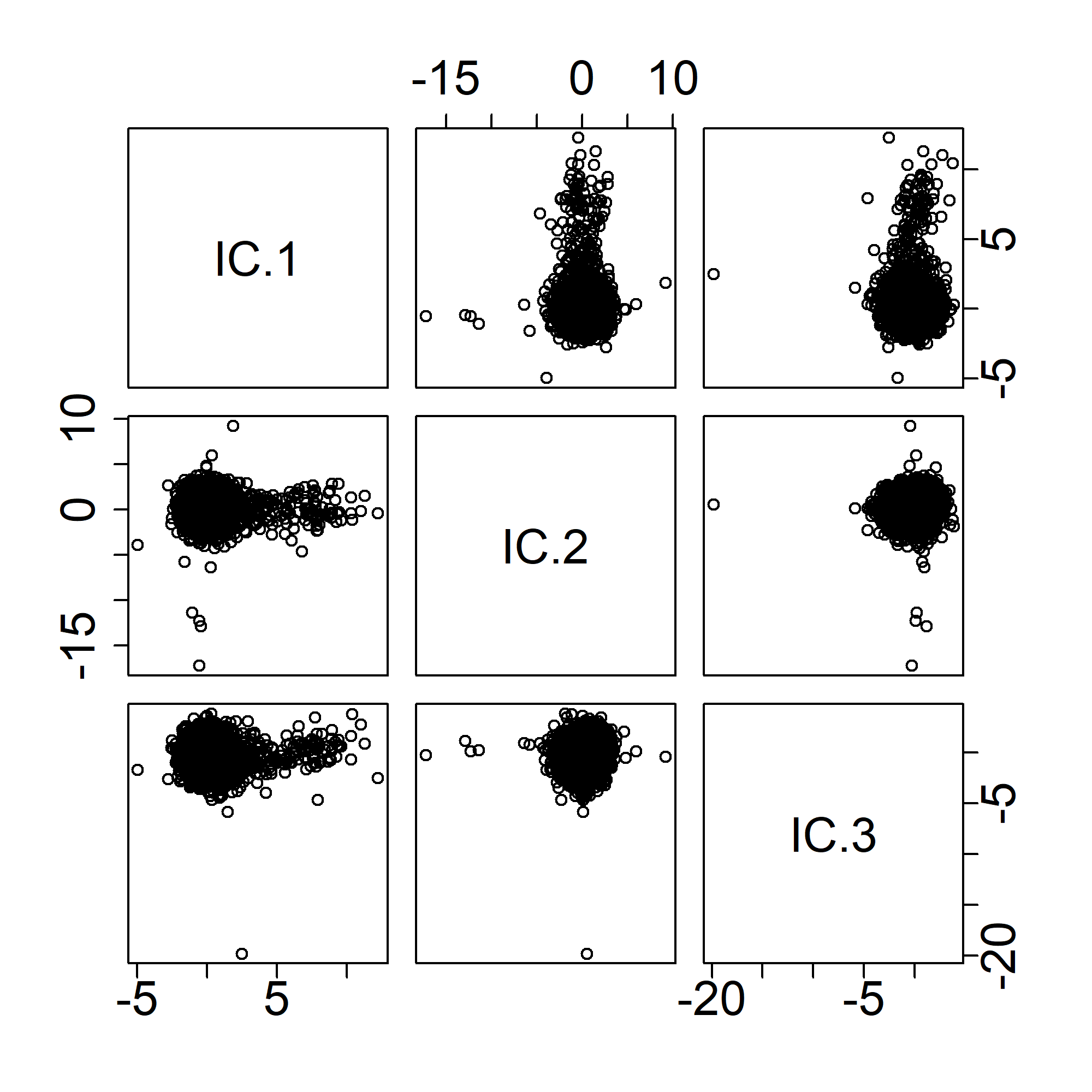}
    \end{minipage}
    \hfill
    \begin{minipage}[t]{0.49\textwidth}
        \centering
        \includegraphics[width=0.98\textwidth]{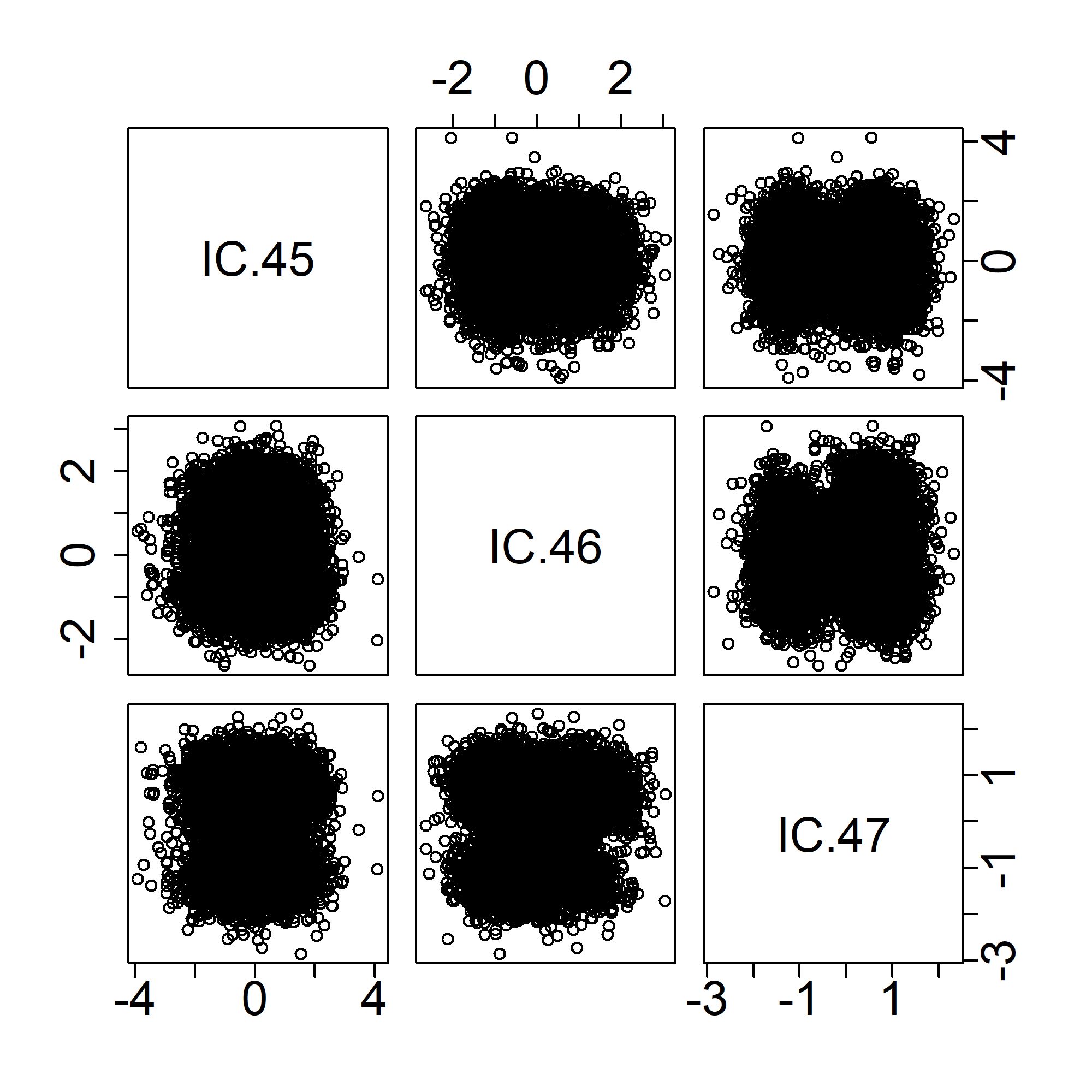}
    \end{minipage}
	\caption{Scatterplots of the first (left) and last (right) three independent components resulting from the compositional FastICA, using adaptive deflation-based FastICA.}
	\label{fig:ica1}
\end{figure}

Due to the kurtosis ordering, the first components show heavy-tailed distributions and they are expected to find outliers or small groupings, while the last components show light-tailed distributions and hence, might find more balanced groupings. Scores of the first and last three independent components are plotted in Figure \ref{fig:ica1}, and the chosen nonlinearities are given in Table \ref{tab:used_non_lin} for all independent components. According to the left plot of Figure \ref{fig:ica1}, one outlier is clearly detected due to its high negative value in the third component (IC.3). According to its loadings, which are collected in Table \ref{tab:FastICALoadings}, IC.3 mostly reflects the relative dominance (with respect to concentrations of all 48 measured metabolites) of phenylalanine (Phe), hexadecanoylcarnitine (C16), octadecenoylcarnitine (C18:1), valine (Val) and hexadecenoyl- and octadecanoylcarnitines in form of C16:1 and C18, respectively, when the higher dominance of the first three metabolites results in decrease of IC.3 and vice versa for the last three stated metabolites. After a deeper investigation of the outlying sample it turned out, that it belongs to a newborn suffering from Phenylketonuria, a metabolic disease which is typically followed by distinctly high absolute blood concentrations of phenylalanine. The measured value was $1\thinspace 014.7$ $\mu\mathrm{mol}/l$, which significantly exceeds the upper norm value set on $120$ $\mu\mathrm{mol}/l$ \cite{wedberg17} and which is represented with the respective high clr value $6.76$. The levels of the remaining metabolites were comparable with the other samples, but particularly the atypical high dominance of Phe over all measured metabolites, which for the rest of samples ranged between $5.72$ and $3.58$ for their clr values, resulted in the high negative value of the third component and therefore clear identification of this non-standard observation.

\begin{figure}
	\begin{minipage}[t]{0.49\textwidth}
		\centering
		\includegraphics[width=0.98\textwidth]{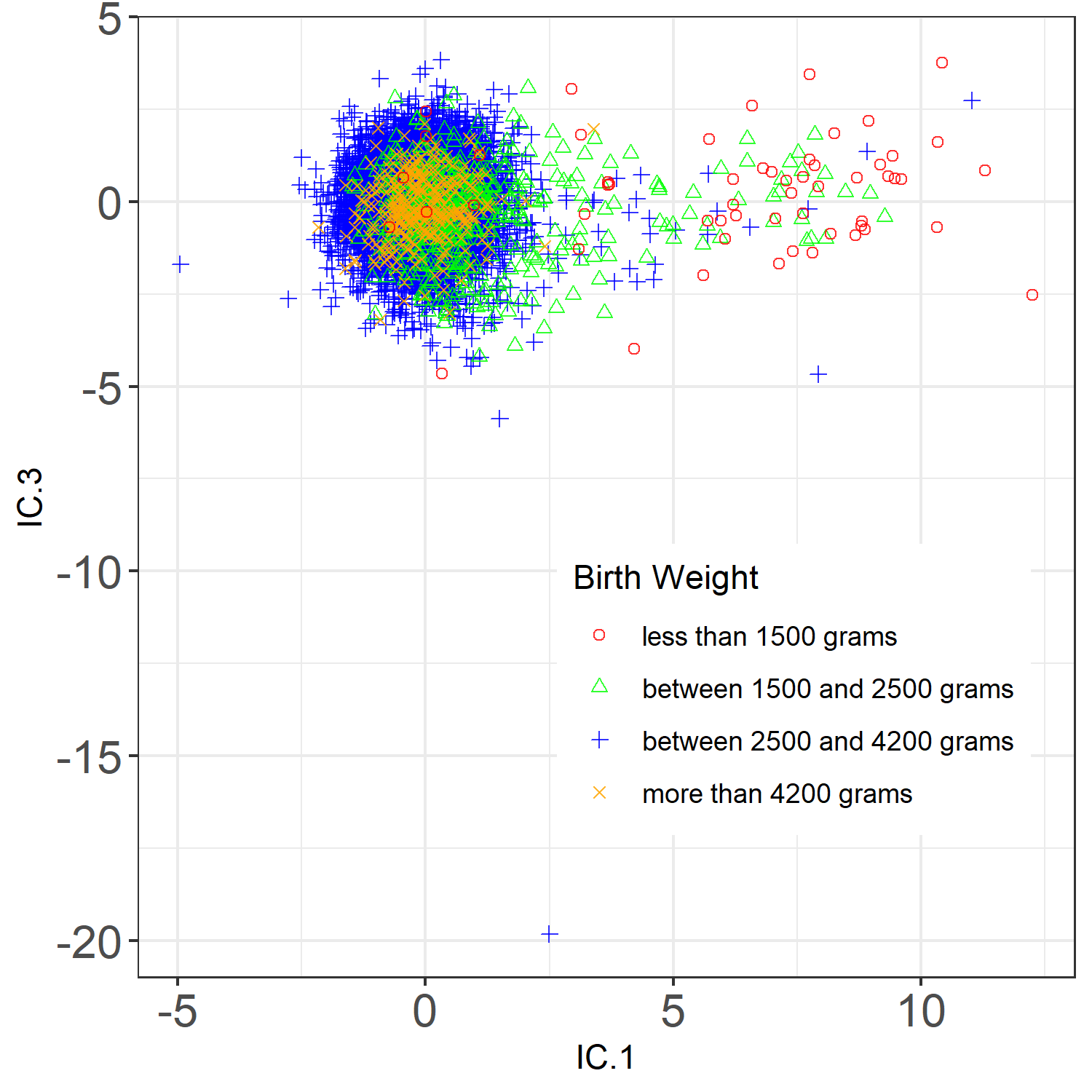}
	\end{minipage}
	\hfill
	\begin{minipage}[t]{0.49\textwidth}
		\centering
		\includegraphics[width=0.98\textwidth]{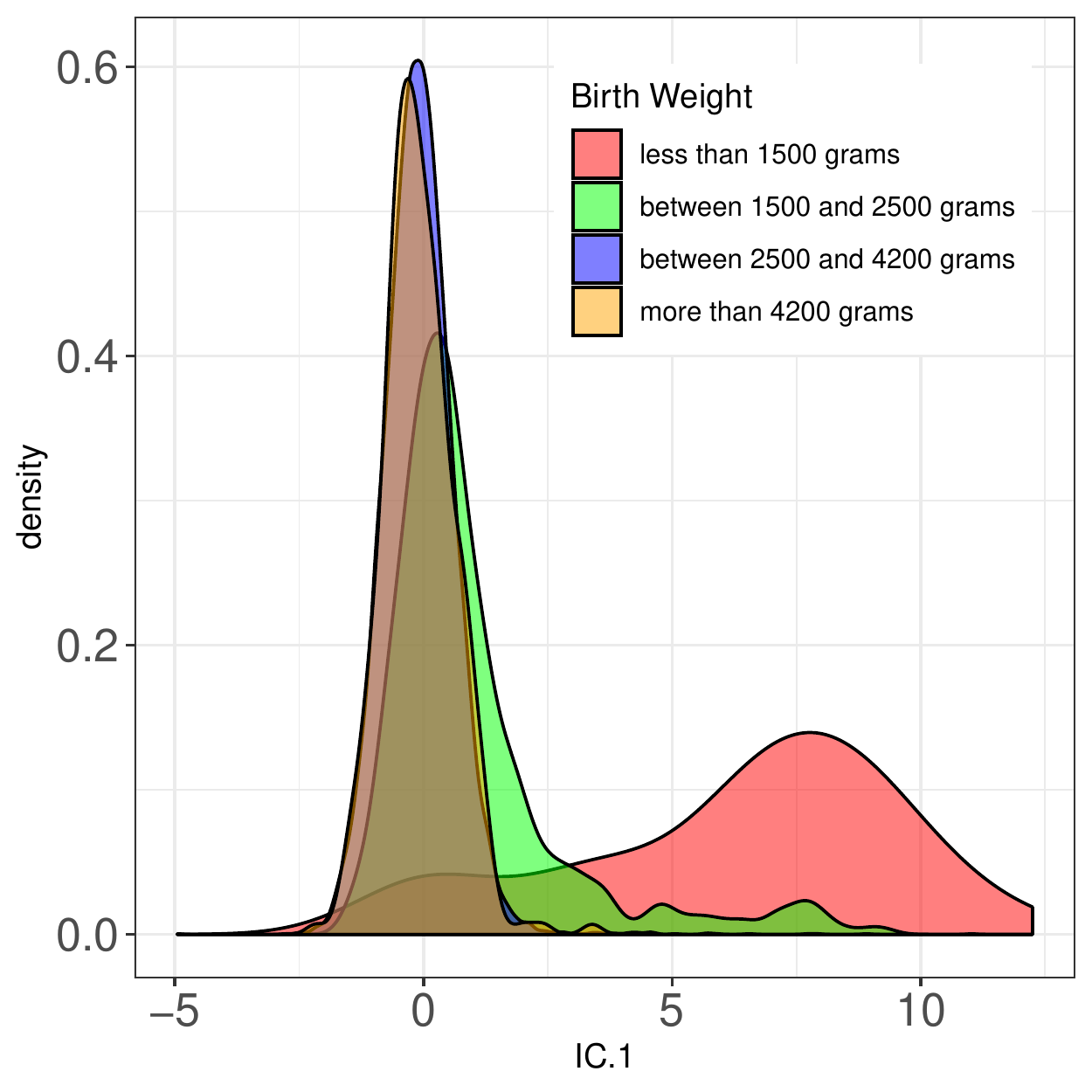}
	\end{minipage}
	\caption{Scatterplots of IC.1 and IC.3 (left) and the kernel density plot of IC.3 (right) with the groups defined according to the birth weight.}
	\label{fig:ica2}
\end{figure}

\begin{figure}
	\centering
	\includegraphics[width=0.88\textwidth]{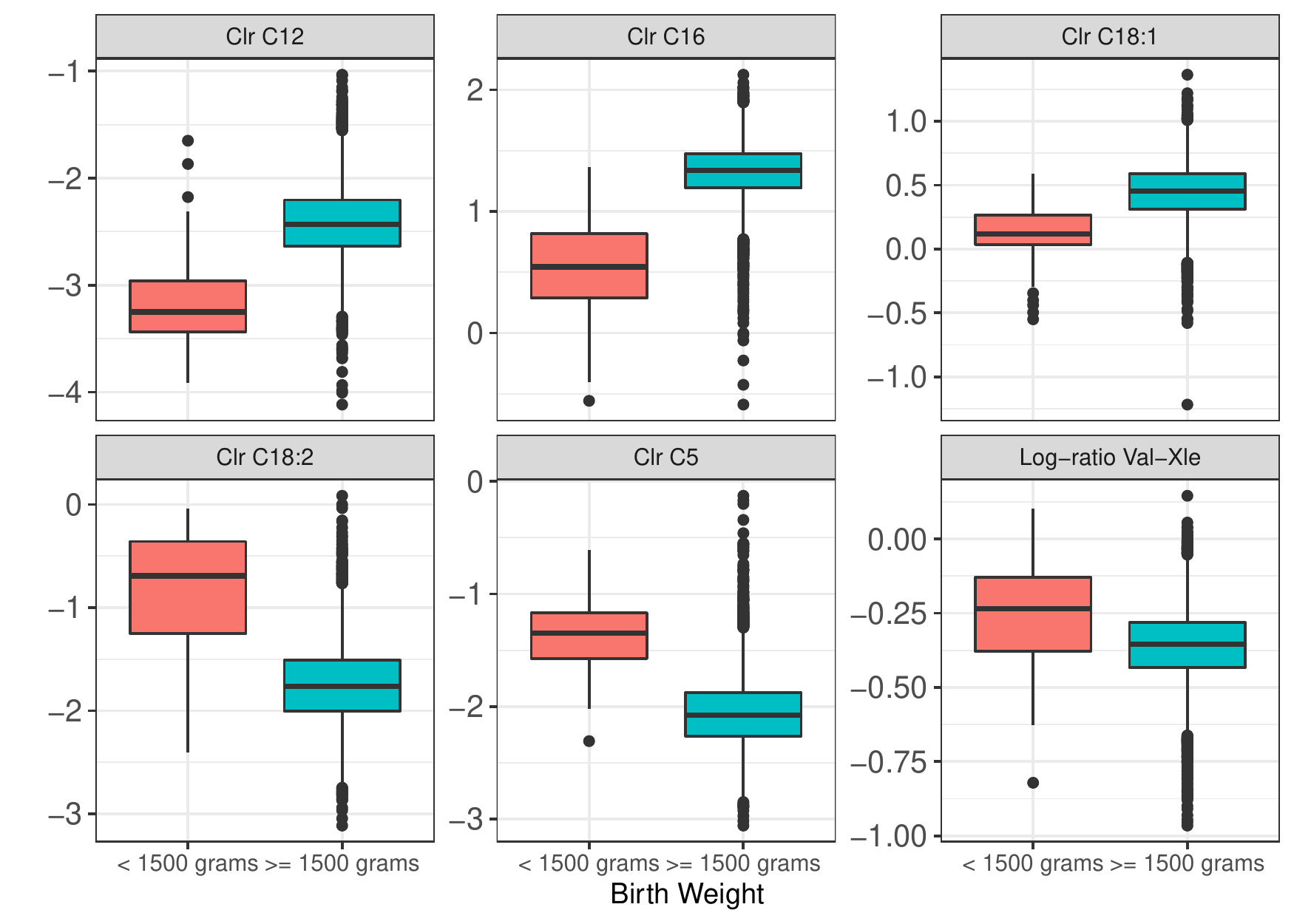}
	\caption{Boxplots of clr as well as log-ratio values of the selected metabolites, which significantly differ for newborns with very low ($< 1500$g) and normal ($>= 1500$g) weight at birth.}
	\label{fig:ica3}
\end{figure}

The next interesting feature is preserved in IC.1. According to Figure \ref{fig:ica1} the values of this component are not very homogeneous across the whole dataset and therefore some specific groups of neonates might be identified. A deeper graphical analysis of the first component (presented in Figure \ref{fig:ica2}) shows that for newborns with a birth weight smaller than 1500 grams higher values of IC.1 are typical. The independent component IC.1 is mostly formed by clr values of acylcarnitines dodecanoylcarnitine (C12), C16 and C18:1, whose high relative dominance over all measured metabolites results in low values of the component and e.g. clr values of  acylcarnitines isovalerylcarnitine/methylbutyrylcarnitine C5 and  linoleoylcarnitine (C18:2) increase the IC.1 values. Even though there are also other metabolites contributing with a high weight to the values of IC.1 (all loadings are collected in Table \ref{tab:FastICALoadings}), the clr values of the selected ones systematically differ for the group of the newborns with low birth weight and therefore these acylcarnitines seem to be responsible for their separation from the remaining neonates. The differences in the selected metabolites are clearly visible in Figure \ref{fig:ica3}. Let us stress here, that the immature neonates  tend to have different diet supplementation, therefore the metabolic profile can substantially differ within this group, but despite of that the proposed ICA method was able to find some similar patterns, detect the important metabolites and separate the low birth weight newborns from the remaining ones. More specifically, artificial nutrition consists of amino acids, lipids, sugars, vitamins etc. Essential unsaturated fatty acids including linoleic acid may be responsible for increased C18:2. The increased blood concentration of the long-chain acylcarnitines (C12, 16, C18:1) as well as of the short-chain C5 carnitine, which then results in high respective clr values, corresponds with previous studies. In \cite{gucciardi2015} the significantly lower amounts of acylcarnitines except the branched-chain acylcarnitines (e.g. C5), that were significantly higher in preterm infants, were described. The latter mentioned are direct products of branched chain amino acid (BCAA) catabolism, therefore its elevate levels may be related to BCAA overfeeding \cite{gucciardi2015, wilson2014}. The difference of several amino acids measured for the premature newborns compared to the others agree with findings in \cite{wilson2014}, where increased levels of several amino acids (arginine, leucine, Orn, Phe, and Val) in the blood spots of premature infants were described. This observation may be related to catabolic state of organisms in these children, amino acid supplementation and immaturity of preterm infants (hepatic maturation, renal insufficiency etc.) \cite{wilson2014, braake2005}. At the first glance counter-intuitively look the opposite loadings by valine (Val) and leucine/isoleucine (Xle), since their levels use to be highly positively correlated. The positive loading by Val and the negative by Xle mean, that the resulting value of IC.1 is affected by the difference of clr values of the respective metabolites, or equivalently by the log-ratio of their measured concentrations, when the higher relative dominance of Val over Xle results in a higher value of IC.1. These findings agree with the data, since slightly higher values of the Val-Xle log-ratio are typical for newborns with a low birth weight (see Figure \ref{fig:ica3}).

\begin{figure}

        \centering
        \includegraphics[width=0.45\textwidth]{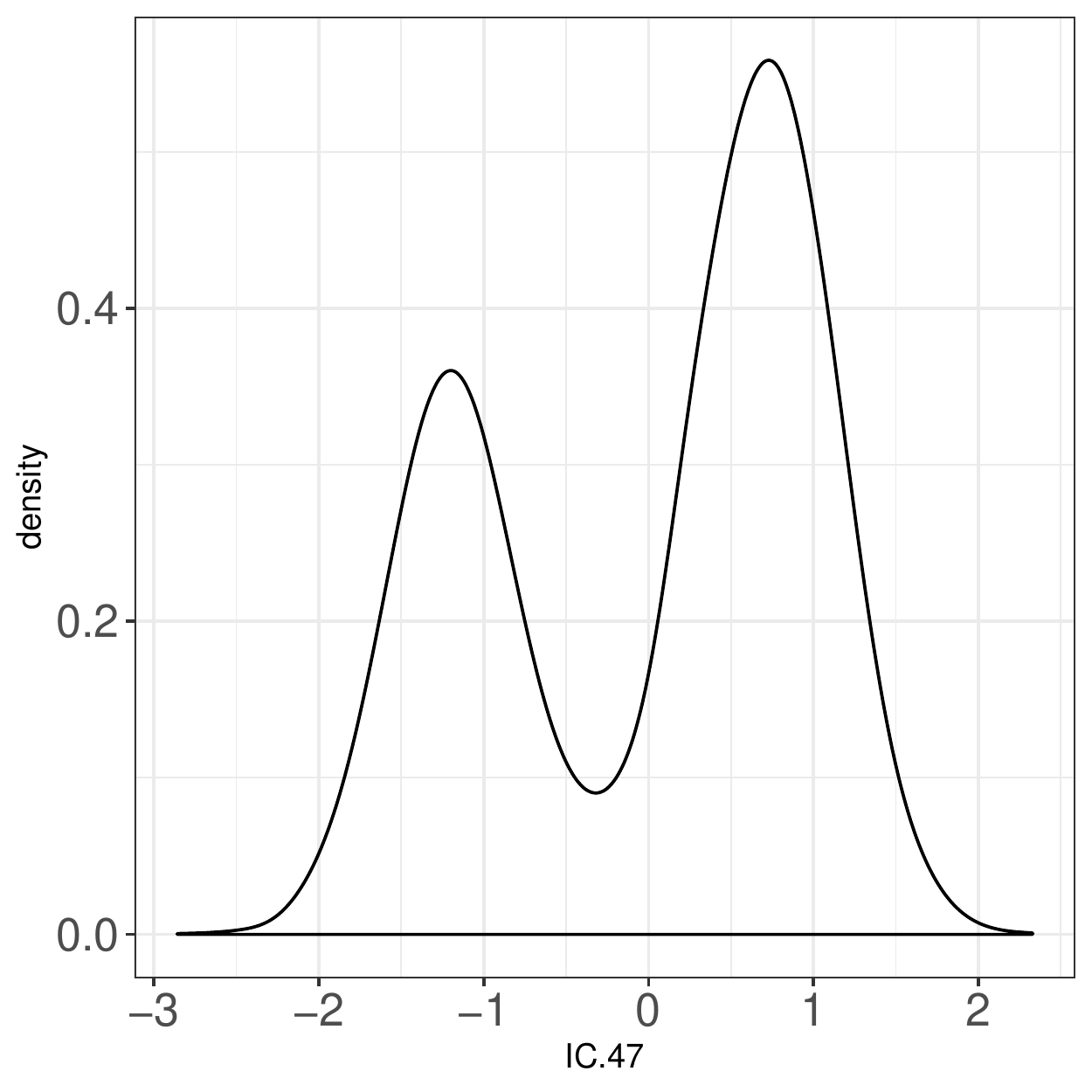}

    \caption{Density plot of IC.47, the bimodal shape shows a clear grouping.}
    \label{fig:ica4}
\end{figure}

An even better visible pattern is formed by the last independent component IC.47, which clearly divides the whole dataset into two groups as seen in Figure \ref{fig:ica4}. According to the loadings (collected in Table \ref{tab:FastICALoadings}), the most contributing are clr values of  metabolites Xle, ornithine (Orn) and lysine (Lys) with a negative effect and methionine (Met), proline (Pro) and valine (Val) with a positive one. The dataset is roughly separated into two groups of observations with values of IC.47 higher and lower than $-0.34$, this value was chosen as the corresponding value of IC.47 at the local minimum in the middle of the density presented in Figure \ref{fig:ica4} (this density was computed with Gaussian kernels and a bandwidth selection with Silverman's rule of thumb). The relative dominance of the six above mentioned metabolites over all measured concentrations do not significantly differ in their values between the two groups, therefore this grouping effect of IC.47 might be hidden in some of their more complex combinations (similarly as in the log-ratio between Val and Xle in the case of weight groups discussed above).

\begin{table}
\caption{Chosen non-linearities $g_i$ for each independent component computed with the adaptive deflation-based FastICA algorithm. Non-linearities are ordered according kurtosis values of the corresponding ICs. In the original ordering IC.44 was the last component thus no non-linearity is given. See Table \ref{NonLin_afica} for the definitions of the functions $g_i$.}
\label{tab:used_non_lin}
\centering
\begin{tabular}{lclclclclc}
  \hline
IC & $g_i$ & IC & $g_i$ & IC & $g_i$ & IC & $g_i$ & IC & $g_i$ \\ 
  \hline
IC.1 & $g_2$ & 		IC.11 & $g_{2}$ &  IC.21 & $g_5$ &  		IC.31 & $g_1$ &	IC.41 & $g_5$ \\ 
IC.2 & $g_2$ &      IC.12 & $g_9$ &    IC.22 & $g_6$ &    IC.32 & $g_{5}$ &    IC.42 & $g_4$ \\ 
IC.3 & $g_6$ &      IC.13 & $g_9$ &    IC.23 & $g_9$ &    IC.33 & $g_{14}$ &       IC.43 & $g_4$ \\ 
IC.4 & $g_2$ &      IC.14 & $g_8$ &    IC.24 & $g_{1}$ &    IC.34 & $g_5$ &    IC.44 & - \\ 
IC.5 & $g_6$ &      IC.15 & $g_6$ &    IC.25 & $g_{1}$ &    IC.35 & $g_{5}$ &    IC.45 & $g_5$ \\ 
IC.6 & $g_8$ &      IC.16 & $g_{10}$ & IC.26 & $g_{4}$ &    IC.36 & $g_5$ &    IC.46 & $g_3$ \\ 
IC.7 & $g_8$ &      IC.17 & $g_{5}$ &  IC.27 & $g_{14}$ &    IC.37 & $g_{12}$ &    IC.47 & $g_6$  \\ 
IC.8 & $g_8$ &      IC.18 & $g_6$ &    IC.28 & $g_{5}$ &    IC.38 & $g_{14}$ &     &  \\ 
IC.9 & $g_9$ &      IC.19 & $g_{10}$ & IC.29 & $g_{10}$ &    IC.39 & $g_4$ & & \\
IC.10 & $g_{8}$ &   IC.20 & $g_6$ &    IC.30 & $g_{8}$ &    IC.40 & $g_{11}$ & & \\

   \hline
\end{tabular}
\end{table}

%

\begin{table}
\caption{The list of loadings for IC.1, IC.3 and IC.47 computed with the adaptive deflation-based FastICA algorithm regarding clr transformed data.}
    \label{tab:FastICALoadings}
\centering
\begin{tabular}{lrrrlrrr}
  \hline
 & IC.1 & IC.3 & IC.47 & & IC.1 & IC.3 & IC.47 \\ 
  \hline
Ala & 0.02 & 0.24 & 0.41 &                C5DC/C6OH & -0.08 & 0.14 & 0.04 \\
  Arg & 0.13 & -0.33 & 0.04 &             C5:1 & -0.05 & 0.11 & 0.13 \\ 
  ArgSucc & 0.00 & -0.09 & -0.01 &        C6 & -0.22 & 0.22 & 0.07 \\ 
  Cit & -0.05 & 0.38 & 0.01 &             C8 & 0.84 & -0.05 & 0.21 \\ 
  Glu & -0.16 & 0.70 & 0.20 &             C8:1 & -0.05 & -0.16 & 0.09 \\ 
  Gly & 0.08 & 1.33 & -0.16 &             C10 & 0.25 & -0.21 & -0.09 \\ 
  His & -0.56 & 0.44 & 0.55 &             C10:1 & -1.05 & -0.01 & -0.21 \\ 
  Lys & 0.20 & 0.38 & -0.46 &             C10:2 & 0.28 & 0.20 & 0.06 \\ 
  Met & 0.53 & 1.03 & 1.01 &              C12 & -1.27 & -0.01 & -0.13 \\ 
  Orn & 0.48 & 0.40 & -0.78 &             C12:1 & -0.09 & 0.10 & -0.09 \\ 
  Phe & 1.02 & -7.30 & -0.40 &            C14 & -0.39 & 0.45 & -0.38 \\ 
  Pro & -1.15 & -0.66 & 0.65 &            C14:1 & 0.88 & -0.46 & 0.04 \\ 
  Thr & -0.17 & -1.71 & -0.01 &           C14:2 & 0.15 & -0.01 & 0.17 \\ 
  Trp & -1.33 & -0.77 & 0.14 &            C14OH & 0.03 & 0.27 & 0.08 \\ 
  Tyr & -0.28 & 1.25 & -0.02 &            C16 & -2.00 & -1.49 & -0.02 \\ 
  Val & 3.09 & 3.86 & 0.59 &              C16:1 & 0.97 & 1.60 & 0.09 \\ 
  Xle & -1.73 & -0.15 & -1.03 &           C16OH & 0.24 & 0.09 & 0.01 \\ 
  C0 & 0.51 & 0.60 & 0.17 &               C16:1OH & 0.05 & 0.01 & 0.24 \\ 
  C2 & 0.08 & -0.30 & -0.25 &             C18 & 2.38 & 1.72 & 0.12 \\ 
  C3 & -0.57 & -0.34 & -0.01 &            C18:1 & -3.46 & -2.64 & 0.00 \\ 
  C3DC/C4OH & 0.29 & -0.23 & 0.34 &       C18:2 & 1.80 & 0.41 & 0.16 \\ 
  C4 & 0.05 & 0.14 & 0.01 &               C18:1OH & -0.00 & 0.27 & -0.15 \\ 
  C4DC/C5OH & 0.07 & 0.49 & -1.42 &       C18:2OH & 0.19 & 0.23 & -0.05 \\ 
  C5 & 0.35 & -0.32 & -0.08 &             C18OH & -0.33 & 0.20 & 0.09 \\ 
   \hline
\end{tabular}
\end{table}

\section{Discussion}
\label{sec:Dis}

In this paper we reviewed some classical independent component analysis methods and showed how these can be applied to compositional data. The key here is that when the ICA methods are affine equivariant it is most natural to use an ilr transformation as the choice of the basis constituting the ilr coordinate system does not matter. For interpretability the link between ilr coordinates and clr coefficients/variables can be easily exploited. This is demonstrated on a metabolomic data set where PCA which is probably the most used multivariate transformation reveals no specific feature on the first few components while ICA reveals several interesting features visible when exploiting the higher order moments information.
Independent component analysis belongs to the larger class of blind source separation methods where for the separation of the latent components often also temporal or spatial information is used. In the context of compositional data such blind source separation methods are for example discussed in \cite{NordhausenOjaFilzmoserReimann2015,NordhausenFischerFilzmoser2019}. But these methods would not be applicable to the metabolomic data set from Section~\ref{sec:CaseStudy} as there is no temporal or spatial information present. The current results, which were discussed mostly in terms of relative dominance of a single compositional part respective to the highest loading of an IC, open new challenges for further research. A more convenient interpretation can be reached e.g. by adaptation of the approach based on principal balances \cite{pawlowsky-glahn11}. However, the loadings of ICs are in general not orthonormal and therefore the principal balances approach is not as straightforward as in the case of PCA. Finally, an extension of the dataset with a group of blood samples collected from neonates with a diagnosed disease can further prove the usefulness of the method.

\section{Acknowledgments}
The work of CM and KN was supported by the Austrian Science Fund (FWF) Grant nr. P31881-N32. The work of KF was supported by The Czech Science Foundation Grant nr. 19-07155S. The work of HJ and AG was supported by The Czech Science Foundation Grant nr. 18-12204S, the grant of Internal Grant Agency, Palack\'y University Olomouc nr. IGA\_LF\_2019\_006 and grant of the Ministry of Health nr. NU20-08-00367.

\bibliographystyle{unsrt}

\begin{thebibliography}{10}

\bibitem{Aitchison1986}
J.~Aitchison.
\newblock {\em The Statistical Analysis of Compositional Data}.
\newblock Chapman \& Hall, London, 1986.

\bibitem{EgozcuePawlowskyGlahn2019}
J.J. Egozcue and V.~Pawlowsky-Glahn.
\newblock Compositional data: the sample space and its structure.
\newblock {\em Test}, 28:599--638, 2019.

\bibitem{facevicova2016}
K.~{Fa\v cevicov\'a}, O.~B\'abek, K.~Hron, and T.~Kumpan.
\newblock Element chemostratigraphy of the devonian/carboniferous boundary - a
  compositional approach.
\newblock {\em Applied Geochemistry}, 75:211--221, 2016.

\bibitem{FilzmoserHronTempl2018}
P.~Filzmoser, K.~Hron, and M.~Templ.
\newblock {\em Applied Compositional Data Analysis}.
\newblock Springer, Cham, 2018.

\bibitem{MoraisThomasAgnanSimioni2018}
J.~Morais, C.~Thomas-Agnan, and M.~Simioni.
\newblock Interpretation of explanatory variables impacts in compositional
  regression models.
\newblock {\em Austrian Journal of Statistics}, 47:1--25, 2018.

\bibitem{pawlowsky2011}
V.~Pawlowsky-Glahn and A.~Buccianti, editors.
\newblock {\em Compositional Data Analysis. Theory and Applications}.
\newblock John Wiley \& Sons, Chichester, 2011.

\bibitem{TrinhMoraisThomasAgnanSimioni2019}
H.~T. Trinh, J.~Morais, C.~Thomas-Agnan, and M.~Simioni.
\newblock Relations between socio-economic factors and nutritional diet in
  {V}ietnam from 2004 to 2014: New insights using compositional data analysis.
\newblock {\em Statistical Methods in Medical Research}, 28:2305--2325, 2019.

\bibitem{comon2010handbook}
P.~Comon and C.~Jutten.
\newblock {\em Handbook of Blind Source Separation: Independent Component
  Analysis and Applications}.
\newblock Academic Press, Amsterdam, 2010.

\bibitem{NordhausenOja2018}
K.~Nordhausen and H.~Oja.
\newblock Independent component analysis: A statistical perspective.
\newblock {\em WIREs: Computational Statistics}, 10:e1440, 2018.

\bibitem{cardoso1989source}
J.-F. Cardoso.
\newblock Source separation using higher order moments.
\newblock In {\em International Conference on Acoustics, Speech, and Signal
  Processing, 1989.}, pages 2109--2112, 1989.

\bibitem{miettinen2014fourth}
J.~Miettinen, S.~Taskinen, K.~Nordhausen, and H.~Oja.
\newblock Fourth moments and independent component analysis.
\newblock {\em Statistical Science}, 30:372--390, 2015.

\bibitem{NordhausenVirta:2019}
K.~Nordhausen and J.~Virta.
\newblock An overview of properties and extensions of {FOBI}.
\newblock {\em Knowledge-Based Systems}, 173:113--116, 2019.

\bibitem{TylerCritchleyDumbgenOja:2009}
D.~Tyler, Critchley, F., L.~D\"umbgen, and H.~Oja.
\newblock Invariant coordinate selection.
\newblock {\em Journal of Royal Statistical Society B}, 71:549--592, 2009.

\bibitem{cardoso1993blind}
J.-F. Cardoso and A.~Souloumiac.
\newblock Blind beamforming for non-gaussian signals.
\newblock In {\em IEEE Proceedings F (Radar and Signal Processing)}, volume
  140, pages 362--370, 1993.

\bibitem{IllnerMiettinenFuchsTaskienNordhausenOjaTheis2015}
K.~Illner, J.~Miettinen, C.~Fuchs, S.~Taskinen, K.~Nordhausen, H.~Oja, and
  F.~J. Theis.
\newblock Model selection using limiting distributions of second-order blind
  source separation algorithms.
\newblock {\em Signal Processing}, 113:95--103, 2015.

\bibitem{clarkson1988remark}
D.~B. Clarkson.
\newblock A least squares version of algorithm {AS} 211: The {FG}
  diagonalization algorithm.
\newblock {\em Journal of the Royal Statistical Society C}, 37:317--321, 1988.

\bibitem{MiettinenNordhausenOjaTaskinen:2013}
J.~Miettinen, K.~Nordhausen, H.~Oja, and S.~Taskinen.
\newblock Fast equivariant {JADE}.
\newblock In {\em IEEE International Conference on Acoustics, Speech and Signal
  Processing (ICASSP) 2013}, pages 6153--6157, May 2013.

\bibitem{VirtaLietzenIlmonenNordhausen2020}
J.~Virta, N.~Lietzen, P.~Ilmonen, and K.~Nordhausen.
\newblock Fast tensorial {JADE}.
\newblock {\em To appear in Scandinavian Journal of Statistics}, 2020.

\bibitem{Hyvarinen99fastica}
A.~Hyv\"arinen.
\newblock Fast and robust fixed-point algorithms for independent component
  analysis.
\newblock {\em IEEE Transactions on Neural Networks}, 10:626--634, 1999.

\bibitem{HyvarinenOja:1997}
A.~Hyv{\"a}rinen and E.~Oja.
\newblock A fast fixed-point algorithm for independent component analyis.
\newblock {\em Neural Computation}, 9:1483--1492, 1997.

\bibitem{Ollila2010}
E.~Ollila.
\newblock The deflation-based {FastICA} estimator: Statistical analysis
  revisited.
\newblock {\em IEEE Transactions on Signal Processing}, 58:1527--1541, 2010.

\bibitem{nordhausen2011deflation}
K.~Nordhausen, P.~Ilmonen, A.~Mandal, H.~Oja, and E.~Ollila.
\newblock Deflation-based {FastICA} reloaded.
\newblock In {\em Proceedings of 19th European Signal Processing Conference},
  pages 1854--1858, 2011.

\bibitem{MiettinenNordhausenOjaTaskinen:2014}
J.~Miettinen, K.~Nordhausen, H.~Oja, and S.~Taskinen.
\newblock Deflation-based {FastICA} with adaptive choices of nonlinearities.
\newblock {\em {IEEE} Transactions on Signal Processing}, 62:5716--5724, 2014.

\bibitem{Hyvarinen:1999}
A.~Hyv\"arinen.
\newblock Gaussian moments for noisy independent component analysis.
\newblock {\em {IEEE} Signal Processing Letters}, 6:145--147, 1999.

\bibitem{Wei2015}
T.~Wei.
\newblock A convergence and asymptotic analysis of the generalized symmetric
  {FastICA} algorithm.
\newblock {\em IEEE Transactions on Signal Processing}, 63:6445--6458, 2015.

\bibitem{MiettinenNordhausenOjaTaskinenVirta:2017}
J.~Miettinen, K.~Nordhausen, H.~Oja, S.~Taskinen, and J.~Virta.
\newblock The squared symmetric {FastICA} estimator.
\newblock {\em Signal Processing}, 131:402--411, 2017.

\bibitem{filzmoser2009}
P.~Filzmoser, K.~Hron, and C.~Reimann.
\newblock Principal component analysis for compositional data with outliers.
\newblock {\em Environmetrics}, 20:621--632, 2009.

\bibitem{hron12}
K.~Hron, P.~Filzmoser, and K.~Thompson.
\newblock {Linear regression with compositional explanatory variables}.
\newblock {\em {Journal of Applied Statistics}}, 39(5):1115--1128, 2012.

\bibitem{egozcue05}
J.~J. Egozcue and V.~Pawlowsky-Glahn.
\newblock Groups of parts and their balances.
\newblock {\em Mathematical Geology}, 37:795--828, 2005.

\bibitem{NordhausenOjaFilzmoserReimann2015}
K.~Nordhausen, H.~Oja, P.~Filzmoser, and C.~Reimann.
\newblock Blind source separation for spatial compositional data.
\newblock {\em Mathematical Geosciences}, 47:753--770, 2015.

\bibitem{pawlowsky-glahn11}
V.~Pawlowsky-Glahn, JJ. Egozcue, and R.~Tolosana-Delgado.
\newblock Principal balances.
\newblock In JJ. Egozcue, R.~Tolosana-Delgado, and MI. Ortego, editors, {\em
  Proceedings of the 4th International Workshop on Compositional Data
  Analysis}, 2011.

\bibitem{Fleischman2013}
A.~Fleischman, J.D. Thompson, and M.~Glass.
\newblock Systematic data collection to inform policy decisions: Integration of
  the region 4 stork (r4s) collaborative newborn screening database to improve
  ms/ms newborn screening in {Washington State}.
\newblock In J.~Zschocke, K.M. Gibson, G.~Brown, E.~Morava, and V.~Peters,
  editors, {\em JIMD Reports - Case and Research Reports, Volume 13}, pages
  15--21. Springer, Berlin, 2013.

\bibitem{kalivodova2015}
A.~Kalivodov\'a, K.~Hron, P.~Filzmoser, L.~Najdekr, H.~{Jane\v ckov\'a}, and
  T.~Adam.
\newblock {PLS-DA} for compositional data with application to metabolomics.
\newblock {\em Journal of Chemometrics}, 29:21--28, 2018.

\bibitem{r_language}
{R Core Team}.
\newblock {\em {R}: A Language and Environment for Statistical Computing}.
\newblock R Foundation for Statistical Computing, Vienna, Austria, 2019.

\bibitem{JADE_package}
J.~Miettinen, K.~Nordhausen, and S.~Taskinen.
\newblock Blind source separation based on joint diagonalization in {R}: The
  packages {JADE} and {BSSasymp}.
\newblock {\em Journal of Statistical Software}, 76(2):1--31, 2017.

\bibitem{fICA_package}
J.~Miettinen, K.~Nordhausen, and S.~Taskinen.
\newblock {fICA: FastICA Algorithms and Their Improved Variants}.
\newblock {\em {The {R} Journal}}, 10:148--158, 2018.

\bibitem{compositions_package}
K.~G. {van den Boogaart}, R.~Tolosana-Delgado, and M.~Bren.
\newblock {\em compositions: Compositional Data Analysis}, 2019.
\newblock {R} package version 1.40-3.

\bibitem{robCompositions_package}
M.~Templ, K.~Hron, and P.~Filzmoser.
\newblock rob{C}ompositions: an {R}-package for robust statistical analysis of
  compositional data.
\newblock In {\em {C}ompositional {D}ata {A}nalysis: {T}heory and
  {A}pplications}, pages 341--355. {J}ohn {W}iley \& {S}ons, 2011.

\bibitem{wedberg17}
A.~van Wegberg, A.~MacDonald, K.~Ahring, A.~Bélanger-Quintana, N.~Blau, A.M.
  Bosch, A.~Burlina, J.~Campistol, F.~Feillet, M.~Giżewska, S.C. Huijbregts,
  S.~Kearney, V.~Leuzzi, F.~Maillot, A.C. Muntau, M.~van Rijn, F.~Trefz, J.H.
  Walter, and F.J. van Spronsen.
\newblock The complete {E}uropean guidelines on phenylketonuria: diagnosis and
  treatment.
\newblock {\em Orphanet Journal of Rare Diseases}, 12:162, 2017.

\bibitem{gucciardi2015}
A.~Gucciardi, P.~Zaramella, I.~Costa, P.~Pirillo, D.~Nardo, M.~Naturale,
  L.~Chiandetti, and G.~Giordano.
\newblock Analysis and interpretation of acylcarnitine profiles in dried blood
  spot and plasma of preterm and full-term newborns.
\newblock {\em Pediatric Research}, 77(1--1):36--47, 2015.

\bibitem{wilson2014}
K.~Wilson, S.~Hawken, R.~Ducharme, B.K. Potter, J.~Little, B.~Thébaud, and
  P.~Chakraborty.
\newblock Metabolomics of prematurity: analysis of patterns of amino acids,
  enzymes, and endocrine markers by categories of gestational age.
\newblock {\em Pediatric Research}, 75:367--373, 2014.

\bibitem{braake2005}
FW. te~Braake, CH. van~den Akker, DJ. Wattimena, JG. Huijmans, and JB. van
  Goudoever.
\newblock Amino acid administration to premature infants directly after birth.
\newblock {\em The Journal of Pediatrics}, 147:457--461, 2005.

\bibitem{NordhausenFischerFilzmoser2019}
K.~Nordhausen, G.~Fischer, and P.~Filzmoser.
\newblock Blind source separation for compositional time series.
\newblock {\em To appear in Mathematical Geosciences}, pages 1--21, 2020.

\end{thebibliography}

\end{document}